\documentclass[final,3p,times]{elsarticle}

\usepackage[english]{babel}
\usepackage{natbib}

\usepackage{amssymb}
\usepackage{amsmath}
 \usepackage{amsthm}
 \usepackage{caption}
\usepackage{subcaption}

\journal{  }

\begin{document}

\begin{frontmatter}


\title{Gravitational Wave Signatures of Warm Dark Matter in Gauge Extensions of the Standard Model }
\author[1]{Lucia A. Popa} 
\ead{lpopa@spacescience.ro}
\address{Institute of Space Sciences (ISS/INFLPR subsidiary),  Atomi\c{s}tilor 409, Magurele-Ilfov, Ro-077125, Romania}

\begin{abstract}
We study the left-right symmetric extension of the Standard Model (LRSM), featuring a
TeV-scale right-handed (RH) gauge boson $W_R$ and three RH neutrinos. 
This setup naturally realises the type-II seesaw mechanism for active neutrino masses. \\
We identify the conditions that yield sufficient entropy dilution to reconcile the keV sterile neutrino dark matter energy density
with observations while inducing an early matter domination (EMD) phase. 
These constrain the lightest active neutrino mass to  $8.59 \times 10^{-10} {\rm eV} < m_{\nu_1} < 5.06 \times 10^{-9} {\rm eV}$. \\
The resulting frequency-dependent suppression of the stochastic gravitational wave (GW) background is set by the mass and lifetime 
of the heavier RH neutrinos. Computing the signal-to-noise ratio (SNR) for future detectors, we find that a blue-tilted 
primordial tensor spectrum can boost the GW signal to detectable levels (SNR > 10) in experiments such as LISA, BBO, and DECIGO.
\end{abstract}

\end{frontmatter}

\section{Introduction}

The gravitational wave (GW) background is one of the most significant predictions of the inflationary paradigm \cite{Grish1975, Staro1979}. These stochastic GWs are generated by quantum fluctuations of the metric, stretched to cosmological scales during the rapid accelerated expansion of the early universe. The
detailed time evolution of the Hubble rate during the expansion determines the transfer
function that describes how the stochastic GWs at different frequencies are red-shifted to
the present day. The resulting GW energy spectrum carries direct information about the 
physics of inflation and the evolution of the universe thereafter. \\
The shape and amplitude of the GW background are primarily determined by the primordial tensor power spectrum, which depends on two key parameters: the tensor initial amplitude $A_t$ and the spectral index  $n_t$ \cite{Turner1993, Turner1997}. 
These parameters are set by the inflationary dynamics and can vary for different inflationary scenarios. 
Furthermore, the spectrum is also affected by the subsequent thermal history of the universe, including the reheating phase and the radiation-dominated (RD) era \cite{Smith2006}. \\
In the standard slow-roll inflationary model, the GW energy spectrum is nearly scale-invariant across a wide range of frequencies. This behavior arises because the relativistic degrees of freedom remain approximately constant for modes re-entering the horizon during the RD era, resulting in minimal distortion of the spectrum. Consequently, the observation of a scale-invariant stochastic GW background would provide compelling evidence for inflation and impose constraints on both the inflationary potential and the reheating dynamics.
Deviations from a scale-invariant GW spectrum can originate from two main sources: the tilt of the primordial tensor power spectrum \cite{Turner1993, White1992, Turner1997}, and modifications in the equation of state of the early universe arising from the interaction properties of elementary particles during and after reheating \cite{Seto2003, Boyle2008}.

One important effect is neutrino free-streaming, first analyzed in Ref.~\cite{Weinberg2004}, 
which leads to a damping of the GW amplitude by approximately 35.5\% in the frequency range $10^{-16}-10^{-10}$ Hz. 
The extension of this analysis to include all Standard Model (SM) particles \cite{Wantanabe2006} demonstrates that
 successive changes in the number of relativistic degrees of freedom leave distinctive imprints on the GW energy spectrum. \\
Moreover, a detailed thermodynamic analysis of SM particles enabled the computation of the temperature dependence of the effective degrees of freedom throughout the expansion history, with results provided in the form of tabulated data and fitting functions \cite{Wantanabe_similar}. In addition, full numerical simulations of the GW spectrum across a wide frequency range, 
incorporating the dynamics of the inflationary scalar field, its decay during reheating, the evolution of relativistic degrees of freedom, and the anisotropic stress from photons and neutrinos, have been performed in Ref.~\cite{Chiba2009}.

The observation of gravitational waves from black hole mergers by the LIGO and Virgo collaborations \cite{LigoVirgo1, LigoVirgo2}, together with GW signals detected by pulsar timing array (PTA) experiments \cite{SKA, EPTA, NANOGrav}, has stimulated the development of various beyond Standard Model (BSM) scenarios for GW production. These include studies on leptogenesis \cite{Giudice1999, Buch2013,Berbig2023}, modified cosmologies and non-thermal dark matter \cite{Nakayama2008, Benal2020, Ghoshal2022}, GWs spectra arising from electroweak phase transitions \cite{FOPT1, FOPT2}, and topological defects  \cite{CS1}. Other works focus on primordial black holes (PBHs) formation and the associated GW signatures \cite{PBH1}, GWs generation by vector and tensor fields \cite{HS1, HS2}, and GWs produced by particle interactions in the early Universe plasma \cite{EU1}. Comprehensive reviews of astrophysical sources and BSM particle physics can be found in Ref. \cite{EU2}, while Ref. \cite{EU3} discusses how current and future GW detectors may discriminate between astrophysical and cosmological BSM signatures in the GWs background. 

An early matter-dominated (EMD) epoch, which can leave imprints on the GW energy spectrum, 
arises naturally in many BSM scenarios in which heavy particles temporarily dominate the energy density 
of the universe before decaying and initiating a radiation-dominated era prior to Big Bang Nucleosynthesis (BBN). \\
Such a phase frequently appears in models with hidden sectors \cite{Dror2016, Berlin2016, Dror2018, Cirelli2019}. 
During the EMD phase, various mechanisms for the production of massive particles can account for the observed 
dark matter (DM) abundance, leading to predictions that are testable through cosmological observations and DM direct 
detection experiments \cite{Giudice1999, Berbig2023, Rouzbeh2022}.

Among the various BSM scenarios for dark matter, a sterile neutrino with a mass in the keV range and a small mixing angle with active neutrinos constitutes a well-motivated Warm Dark Matter (WDM) candidate \cite{sn1, sn3, sn4}.\\
Emerging from a minimal extension of the Standard Model, the keV-scale sterile neutrino can simultaneously explain active neutrino oscillations, the observed dark matter abundance, and the matter-antimatter asymmetry of the universe \cite{sdm1, sdm2}.\\
Several production mechanisms for keV-scale sterile neutrinos have been proposed. 
One of the earliest is the Dodelson-Widrow (DW) scenario \cite{DW}, in which sterile neutrinos are produced via non-resonant oscillations with active neutrinos (NRP). 
However, this mechanism is now ruled out by structure formation constraints, as it produces sterile neutrino velocity
 distributions that are too hot \cite{NRP1, NRP2}. \\
 An alternative is resonant production (RP), known as the Shi-Fuller mechanism \cite{SF}, in which a large lepton asymmetry enables efficient active-to-sterile neutrino conversion \cite{RP5}. In this scenario, the sterile neutrino parameters required to reproduce the observed dark matter abundance are broadly consistent with current cosmological observations, including constraints from the Local Group and high-redshift galaxy counts \cite{RP5}. Nonetheless, some tension remains with Lyman-$\alpha$ forest data \cite{Merle1}. \\
Additionally, keV sterile neutrino dark matter production via particle decays has been extensively discussed (see Ref.~\cite{WhitePaper} and references therein). 

The lower mass limit of sterile neutrino dark matter is constrained by the universal Tremaine-Gunn bound \cite{Tremain}, which applies when all dark matter is composed of sterile neutrinos. A more stringent constraint arises from analyses of the dark matter phase-space distribution in dwarf spheroidal galaxies, leading to  $m_{N_1}>1.8$ keV for sterile neutrino non-resonant production (NRP) \cite{Boya_ll}. 
This bound was revisited in  Refs. \cite{Les3} in the frame of  $\Lambda$Cold+Warm Dark Matter ($\Lambda$CWDM) model where WDM in the form of sterile neutrinos represents a fraction, $f_{DM}$, from total DM.\\
From a combined analysis of WMAP5 and Lyman$-\alpha$ datasets it was found that  
if $f_{DM}=1$ (pure $\Lambda$WDM model)  the lower bound of sterile neutrino mass in NRP case 
is $M_{N_1} >1.6$ keV (at 95\%CL),  that scales as $f^{1/3}_{DM}$ \cite{Slosar}.
The same analysis show that, below a certain threshold, $f_{DM}$ approach a constant value
because the contribution of WDM component become too small to be constrained by the data.  \\
The combine analysis of the CMB anisotropy and lensing data, cosmic shear observations,
and low-redshift BAO measurements \cite{Popa2021}, yields a sterile neutrino mass
$M_{N_1} = 7.88 \pm 0.73 $ keV (68\%CL) and $f_{DM} = 0.86 \pm 0.07$ (68\% CL) 
consistent with Lyman-$\alpha$ forest constraints that exclude $f_{\rm DM}=1$ at $3\sigma$ \cite{Slosar}. 
A review of sterile neutrinos as potential dark matter candidate can be found in Refs.~\cite{Aba2017,Boyarsky2019}.

Sterile neutrino dark matter is unstable. To serve as a viable dark matter candidate, 
its lifetime $\tau_{N_1}$  must exceed the age of the universe, 
$\tau_u \sim 10^{17}$ s. \\
A more stringent constraint on $\tau_{N_1}$ arises from its radiative decay channel  $N_1 \rightarrow \nu \gamma$ 
in which  DM sterile neutrino decays into an active neutrino and a photon of 
energy $E_\gamma~=~M_{N_1}/2$. This photon lies within the sensitivity range of current and upcoming X-ray observatories
\cite{Jeltema,Riemer,Loew,Urban,Malyshev}.  
The decay width of this process,
imposes an upper bound on the active-sterile neutrino mixing angle $\theta^2_1 \le  1.8 \times 10^{-5} ({\rm 1\, keV}/M_{N_1})^5$,
leading to  the DM sterile neutrino lifetime  $\tau_{N_1} \sim 10^{24}$ s, six times longer than the age of the universe \cite{sn3,sdm1}. \\ 
For $M_{N_1} \ge 2 $keV  the resulting contribution of DM sterile neutrino $N_1$ to the active neutrino masses, 
$ \delta m_{\nu} \sim \theta^2_1 M_{N_1}$, 
remains below the uncertainty in the solar neutrino mass-squared difference, 
as indicated by global fits to neutrino oscillation data \cite{Global_fit}. 
Consequently, two additional right-handed (RH) neutrinos
remains below the uncertainty in the solar neutrino mass-squared difference, as indicated by global fits to neutrino oscillation data \cite{Global_fit}. consequently two additional right-handed (RH) neutrinos, 
$N_2$ and $N_3$  are required to  reproduce  the observed neutrino oscillation pattern. 

In this paper, we assume an inflationary reheating scenario consistent with the normal ordering (NO) of right-handed (RH) neutrino masses
$M_{N_1}< M_{N_2} < M_{N_3}$. We further 
assume that DM sterile neutrino $N_1$ is thermally produced as a relativistic particle via freeze-out, while $N_2$ and $N3$ 
decouple while relativistic and decay out of equilibrium after the freeze-out of $N_1$. 

 This scenario can be realised within the left-right symmetric extension of the Standard Model (LRSM) 
 which introduces a right-handed charged gauge boson $W_R$ with a mass at the TeV scale, 
 and employs  the type-II seesaw mechanism to generate active neutrino masses \cite{Zhang2008,Maiezza2010,Magg1980}. 
 In this model, the mass spectrum of the RH neutrinos  leads to the same hierarchical
active neutrino spectrum.  The presence of $W_R$ boson  plays  a key role in accurately predicting the dark matter relic abundance, ensuring 
that,  for RH neutrinos  that decouple while relativistic, 
their freeze out temperatures, and consequently their yields, are expected to coincide to a very good approximation \cite{Berzukov2010}.\\
Once the heavier RH neutrinos become non-relativistic, they  behave as matter and, if are sufficiently long-lived, can dominate the energy density of the universe. This results in an epoch of early matter domination (EMD), which ends with their decays and the associated release of a substantial amount of entropy.\\
We show that this entropy injection simultaneously dilutes the abundance of the DM sterile neutrino  $N_1$ bringing  it
into agreement with observational constraints \cite{sn3, Berzukov2010}, 
and suppresses the GW energy density spectrum at scales that re-enter the horizon prior to, or during the decay of the heavier RH neutrinos, 
leaving measurable imprints on the spectral shape of the gravitational wave background \cite{Giudice1999, Berbig2023, Allahverdi2022}.

The paper is organised as follows. Section 2 presents the constraints and requirements for DM sterile neutrino production. 
Section 3 analyses  the propagation of inflationary tensor perturbations
as gravitational waves.  Section 4  discusses the imprints left on 
the gravitational waves background spectrum by the 
the RH neutrino  decay, 
and evaluate the capability of various GWs experiments in searching 
these specific  signatures in terms of signal-to-noise ratio. 
Our conclusions are summarised in Section 5.

\section{Constraints and requirements for DM sterile neutrino production} 

\subsection{DM sterile neutrino abundance}

Inclusion of the RH charged gauge boson $W_R$ 
with TeV  mass   introduces new annihilation channels and modifies the freeze-out dynamics of RH neutrinos.
The scattering of RH neutrinos with the light SM fermions, mediated by the heavy
gauge bosons keep them in thermal equilibrium. Their freeze out (decoupling) temperature $T_f$ can be estimated from the out-of-equilibrium condition  
$\Gamma_{N}=H(T_f)$, where   $\Gamma_{N}$ is the interaction rate of and $H(T_f)$
is the Hubble expansion rate in the radiation dominated universe  \cite{Berzukov2010,Neme2012,Boyarsky2019} :
\begin{eqnarray}
\Gamma_{N}(T_f) \simeq G^2_F T^5_f  \left(  \frac{M_W}{M_{W_R}} \right)^4\,, 
\hspace{1cm}
H(T_f)=\sqrt{    \frac{4 \pi g_{*f} }  {45} } \frac{T^2_f }{M_{pl}}    \,,
\end{eqnarray}
where  $g_{*_f}$  counts the number of relativistic degrees of freedom at $T_f$,  
$G_F=1.66 \times 10^{-5}$ GeV$^{-2}$ is the Fermi constant, $M_W=80.34$ GeV is the mass of W boson \cite{PDG} and 
$M_{pl}=1.2 \times 10^{19}$ GeV is the Planck mass.
The out-of-equilibrium condition leads to  \cite{Neme2012}:
\begin{eqnarray}
\label{Tf}
T_f\simeq 4\, {\rm MeV}\, \left(  \frac{g_{*f}}{10.75}\right)^{1/6} \left( \frac{M_{WR}}{M_W}\right)^{4/3}\,.
\end{eqnarray}
For RH neutrinos  which freezes out while relativistic, 
the number density per entropy density (the yield) at $T_f$ is:
\begin{equation}
\label{abund}
Y (T_f)= \frac{n_{N_1} (T_f)}  {s(T_f)}\simeq \frac{1}{g_{*f}}\frac{123 \zeta(3)}{4 \pi^4}       \,.
\end{equation}
If the gauge interactions of RH neutrinos are universal, their freeze out temperatures, and consequently their yields, are expected to coincide.\\

The contribution of keV sterile neutrino $N_1$ with present-day energy density $\Omega_s$, to the present-day
 total dark matter energy density $\Omega_{DM}$,
is observationally constrained to  energy density fraction $f_{DM} \leq 1$
Accordingly, using Eq. (\ref{abund}) and taking into account that  the yields of RH neutrinos are  thermally conserved quantities,
 $f_{DM}$ can be written as:
\begin{eqnarray}
\label{f_DM}
f_{DM}=\frac{\Omega_s}{\Omega_{DM}} = \frac{Y_{N_1}  s_0}{\Omega_{DM} \rho_c}
\, \left( \frac{M_{N_1}}{1\,{\rm keV} }\right) \, \frac{1}{\Delta_S} \,,
\end{eqnarray}
where: $M_{N_1}$ is the sterile neutrino mass, $s_0=2889.2$ cm$^{-3}$ is the present entropy density and 
$\rho_c=1.05368 \times10^{-5}$ h$^2$ GeV cm$^{-3}$ is the critical energy density.  $\Omega_{DM}$ is constrained 
by the cosmological observations to $\Omega_{DM}=0.228 \pm 0.005$ at $3\, \sigma$ \cite{Planck_par}. \\
In the following  we adopt the sterile neutrino masses $M_{N_1}$ in the range $1.6 \div 8$ keV, consistent with the observational mass bounds of keV sterile neutrino \cite{Boyarsky2019}.\\
The dilution factor $\Delta_s$ in Eq. (\ref{f_DM}) accounts for the post freeze-out entropy injection needed
 to satisfy the observational constraint $f_{DM} \leq 1$. 
The value of $\Delta_s$ required to avoid the over closure  of the of the universe 
is then given by:
\begin{equation}
\label{dilut_N1}
\Delta_s = \frac { Y_{N_1} s_0} { \Omega_{DM} \rho_c} \left( \frac{M_{N_1}}{1 \, {\rm keV} }\right) \,.
\end{equation}

\subsection{Early matter domination (EMD) onset and the entropy dilution}

The large entropy injection required to achieve the correct abundance of the sterile neutrino 
$N_1$ can be generated through the decay of a heavy, long-lived particle with a relatively short lifetime. 
In the LRSM, the only particles capable of such late 
decays with a significant impact on entropy production are the RH neutrinos $N_2$ and $N_3$.

For simplicity, in what follows we focus on $N_2$ as the diluter. 
In this scenario, the RH neutrino $N_2$
 reaches thermal equilibrium at early times through its interactions with the thermal bath. After becoming non-relativistic, it behaves as a matter component and, if sufficiently long-lived, can come to dominate the energy density of the universe. \\
 This leads to a period of early matter domination (EMD), which ends once  $N_2$ decays, 
 restoring a radiation-dominated (RD) universe and setting the conditions for the Big Bang Nucleosynthesis (BBN). 

\begin{figure}
\begin{center}
\includegraphics[height=6cm]{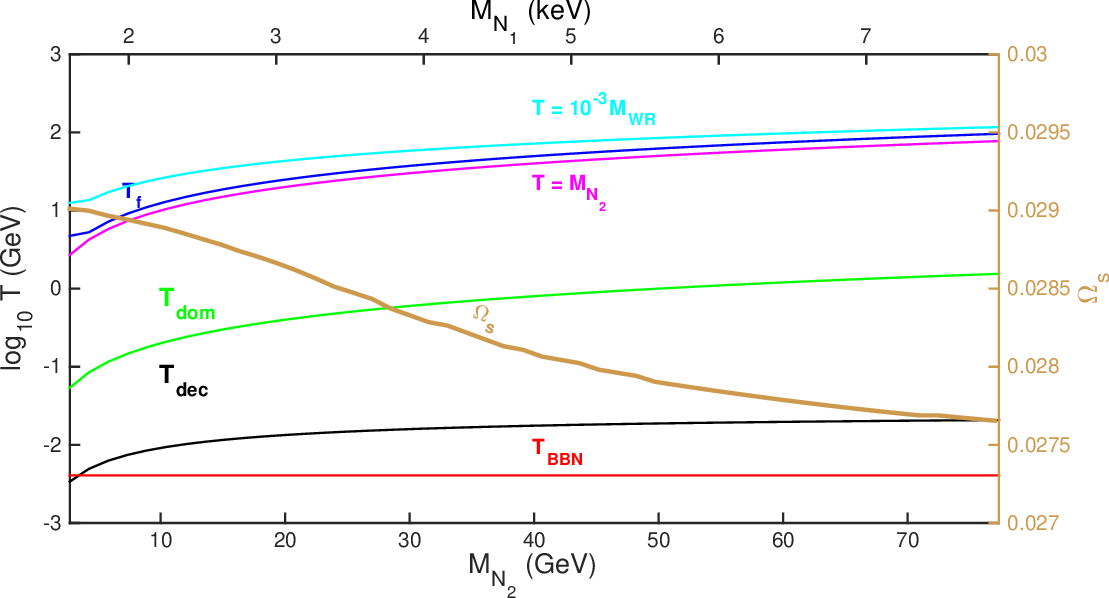}
\caption{ Evolution of the temperature in the $M_{N_1} - M_{N_2}$ parameter space across different 
stages, after imposing the conditions given in
Eqs. (\ref{req_Tf}), (\ref{MWR}) and (\ref{gamma_N2}). The $N_1$ keV sterile neutrino  energy density  
$\Omega_s$ is also presented. \label{fig1} } 
\end{center}
\end{figure}

The important stages of the thermal history in this scenario and their requirements are as follows: 
\begin{itemize}
\item{ In order to achieve enough  dilution,  $N_2$ neutrino  freeze out temperature  $T_f$ should exceeds its mass, 
otherwise the yield in Eq. (\ref{abund}) receives a suppression factor $e^{-M_{N_2}/T_f}$:
\begin{equation}
\label{req_Tf}
T_{f}> M_{N_2}\,.
\end{equation}
As $N_2$  freeze out at temperature $T_f$ as given by Eq. (\ref{Tf}),
 the constraint (\ref{req_Tf})  leads to the bound on the RH gauge boson mass $M_{WR}$:
\begin{equation}
\label{MWR}
M_{WR} \ge 7 \, {\rm TeV}   \frac{1}{g_{*f}^{1/8} } \left( \frac{M_{N_2} }{1\, {\rm GeV} } \right) ^{3/4}\,.
\end{equation}
}
\item{The transition of $N_2$ to the non-relativistic regime is around at temperature $T=M_{N_2}$.
At this stage  the $N_2$ energy density redshifts more slowly than
that of radiation, scaling as $\rho_{N_2} \propto a^{-3}$ compared to $\rho_r \propto a^{-4}$,
and $N_2$ can come to dominate the total energy density of the universe. 
We denote the temperature at which this transition occurs by $T_{dom}$, and the corresponding Hubble expansion rate by
$H_{dom}$. The necessary condition for ${N_2}$ dominance is:
\begin{equation}
\label{req_Hdom}
\Gamma_{N_2}<H_{dom}\,.
\end{equation}
The condition $\rho_{N_2} \simeq \rho_r$ can be used to calculate $H_{dom}$ and $T_{dom}$ \cite{Rouzbeh2022}:
\begin{eqnarray}
\label{req1_Hdom}
H_{dom}= \frac {g^{2/3}_{*dom}} {g^{8/3}_N} H(T=M_{N_2})\,,
\hspace{1cm} T_{dom}=\frac{7}{4} \frac{M_{N_2}}{g_{*}(T_{dec})} \simeq 2 \% M_{N_2}\,.
\end{eqnarray}
Here $g_{*dom}$  and $g_{*}(T_{dec})$ are the number of relativistic
degrees of freedom at $T_{dom}$ and $T_{dec}$ while 
$g_{*N}=2$ accounts for the two degrees of freedom associated with  $N_2$ neutrino.\\
}
\item{Before decaying, $N_2$ becomes non-relativistic as long as its total decay width is smaller than the Hubble rate at temperatures around $M_{N_2}$, i.e. $\Gamma_{N_2} \ll H(M_{N_2})$.
Assuming that the decay is instantaneous and the decay products thermalise quicly with the radiation bath, 
the decay temperature $T_{dec}$ can be estimated from the condition $\Gamma_{N_2}~=~H(T_{dec})$ as:
\begin{eqnarray}
\label{Tdec}
T_{dec} \simeq \left( \frac{90}{8 \pi^3 g_{*} (T_{dec})} \right)^{1/4} 
\sqrt{\Gamma_{N_2}M_{pl}}
\end{eqnarray}
}
\item{The EMD era ends once  $N_2$ decays and mast be  completed before the onset of
the BBN. This constraint imposes a lower bound on the decay temperature, typically $T_{dec} > T_{BBN}=4$ MeV, corresponding to:
\begin{eqnarray}
\label{BBN_constrain}
\Gamma_{N_2} > H_{BBN} \sim 10\, s^{-1}\,,
\end{eqnarray}
leading to $N_2$ lifetime bound $\tau_{N_2} =\Gamma^{-1}_{N_2} \sim 1 \, s$.
}
\item{After  $N_2$ decay is completed, EMD ends and the universe is reheated. 
The entropy injection associated with the decay products dilutes the pre-existing relics. 
This dilution is quantified by the entropy dilution factor $\Delta_s$ defined as the ratio  entropy density before and after
 $N_2$ decay \cite{Giudice1999,Rouzbeh2022, Berbig2023}:
 \begin{eqnarray}
 \Delta_s \equiv  \frac{s_{after}} {s_{before}}\,
 \end{eqnarray}   
where $s_{before}$ is the entropy density of the radiation existing from prior stages at $H=\Gamma_{N_2}$ 
while $s_{after}$ is the entropy density generated by $N_2$ decay. \\
Using the energy density conservation:
\begin{eqnarray}
M_{N_2}Y_{N_2} s_{before}=\frac{3}{4}s_{after} T_{dec} \,,
\end{eqnarray}
the dilution factor is obtained as:
\begin{eqnarray}
\label{dilut_N2}
\Delta_s \simeq 1.8 g_f ^{1/4}  
\frac{Y_{N_2} M_{N_2}} { \sqrt{\Gamma_{N_2} M_{pl}} } \,.
\end{eqnarray}
It is worth noting that, in this scenario, the freeze-out temperatures of the DM sterile neutrino $N_1$ coincides with that of 
$N_2$. \\
Combining Eqs. (\ref{dilut_N1}) and (\ref{dilut_N2})} we obtain the $N_2$ decay width:
\begin{equation}
\label{gamma_N2}
\Gamma_{N_2} \simeq 0.38 \times 10^{-6}g^{-1/2}_{f} \frac{M^2_{N_2}}{M_{pl}}\,
\left(  \frac{1\, {\rm keV}} {M_{N_1}} \right)^2 \,.
\end{equation}
Eq. (\ref{gamma_N2}) provides the proper entropy dilution if $N_2$ decouple while still relativistic as indicated in (\ref{req_Tf}) and satisfy the BBN constraint (\ref{BBN_constrain}). These constraints translate into a bound on the mass of $N_2$:
\begin{eqnarray}
\label{M_N2}
M_{N_2} > \left( \frac{M_1}{1 \,{\rm keV}} \right)  (1.7 \div 10) {\rm GeV}\,.
\end{eqnarray} 
\end{itemize}
Figure~\ref{fig1} presents the temperature evolution in $M_{N_1} - M_{N_2}$ parameter space 
across different stages  discussed above,  after imposing the conditions in Eqs. ({\ref{req_Tf}}), (\ref{MWR}) and (\ref{gamma_N2}).
The $N_1$ energy density $\Omega_sh^2$ is also presented. 
For reference, this should compared with the present energy density of active neutrinos $\Omega_{\nu} \simeq 0.0011$. \\
In Figure~\ref{fig2} we show  the conditions for early matter domination (EMD) given in Eqs. (\ref{req_Hdom}) and
(\ref{BBN_constrain}). 

In the framework of the left-right symmetric model (LRSM) where the active neutrino masses
are generated by the type II seesaw mechanism,  the mass spectrum of RH neutrinos leads to the
 same hierarchical spectrum 
for the active neutrinos, implying  that the ratio of the mass eigenvalues
of active and sterile neutrinos are identical \cite{Giudice1999,Berzukov2010}. 
For the normal mass ordering, this relation can be expressed as:
\begin{equation}
\frac{m_{\nu_1}}{m_{\nu_2} }= \frac{M_{N_1}}{M_{N_2}}\,,\hspace{0.5cm}\frac{m_{\nu_2}}{m_{\nu_3} }= \frac{M_{N_2}}{M_{N_3}} \,.
\end{equation}
Consequently, the parameter space  $M_{N_1} - M_{N_2}$ imposes  bounds  on the active neutrino mass $m_{\nu_1}$
and on the RH neutrino mass $M_{N_3}$.
For the normal ordering,  $m_{\nu_2}=\sqrt{\Delta^2 m_{sol.}}\simeq 8.6 \times 10^{-3}$eV
and $m_{\nu_3}=\sqrt{\Delta^2 m_{sol.}+\Delta^2 m_{atm}} \simeq 0.05 $eV where $\Delta^2 m_{sol.}=7.420^{+0.21}_{-0.20} \times 10^{-5} {\rm eV}$ 
and 
$\Delta^2 m_{atm.}=2.517^{+0.026}_{-0.028} \times 10^{-3} {\rm eV}$ 
are the solar and atmospheric neutrino mass-squared differences respectively obtained from 
the global fit to neutrino oscillation data \cite{Global_fit}, we obtains: 
\begin{eqnarray}
8.59 \times 10^{-10} \,{\rm eV} \,& < &m_{\nu_1}  < 5.06 \times 10^{-9} \,{\rm eV}\,   \\
15.21\,{\rm GeV}\,& < & M_{N_3} \, < 450.57  \,{\rm GeV}\,.
\end{eqnarray}  

\begin{figure}
\begin{center}
\includegraphics[width=6cm]{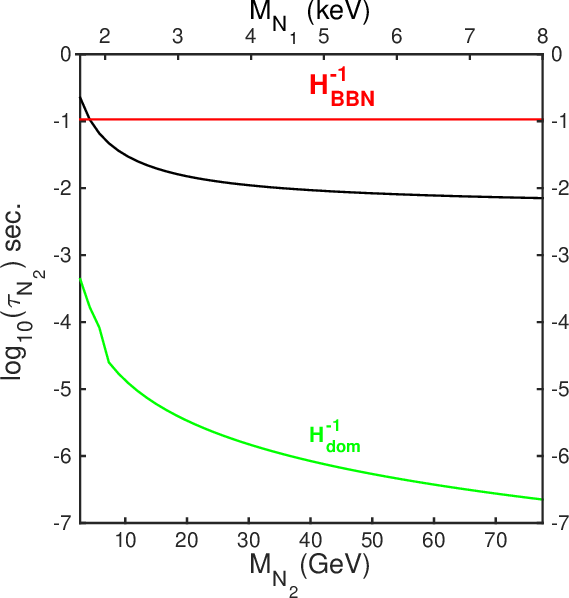}
\caption{ The conditions for the early matter domination (EMD) from Eqs. (\ref{req_Hdom}) and (\ref{BBN_constrain})  presented in 
$M_{N_1} - M_{N_2}$ parameter space. \label{fig2}
Here $\tau_{N_2}=\Gamma^{-1}_{N_2}$  denotes the lifetime of  sterile neutrino $N_2$  
while $H^{-1}_{dom}$   and $H^{-1}_{BBN}$  correspond to the lifetimes of the EMD phase 
and of the Big Bang Nucleosynthesis (BBN), respectively.  \label{fig2} }
\end{center}
\end{figure} 

\section{The spectrum of the gravitational waves}

The gravitational waves (GWs) energy density spectrum today  can be written as \cite{Berbig2023,Watanabe2006,Caprini2018}:
\begin{eqnarray}
\label{OmegaGW}
\Omega_{GW}(k)=\frac{1}{12 H^2_0 a^2_0} [ T^{'}(k,\tau_0) ]^2{\mathcal P}(k)\,,
\end{eqnarray}
where $k$ is the GW wave number, $\tau_0=2/H_0$ is the conformal time today,  
${\mathcal P}(k) $  is the primordial power spectrum of the tensor modes, $[ T^{'}(k,\tau_0) ]^2$ is the 
derivative of the transfer  function with respect to the conformal time, $a_0=1$ is the present-day scale factor and 
$H_0 \simeq 2.2 \times 10^{-4}$ Mpc$^{-1}$ is the present Hubble expansion rate \cite{Planck_par}. \\
In the sub-horizon regime relevant in this paper, i.e. $k\tau_0 \gg 1$, one typically uses the approximation $[ T^{'}(k,\tau_0) ]^2 \sim k^2T^2(k,\tau_0)$ 
leading to:
\begin{equation}
\label{omega_gw}
\Omega_{GW}(k)= \frac{1}{12}  \left( \frac{k} {a_0 H_0}\right)^2 T^2(k){\mathcal P}(k)\,,
\end{equation}

The primordial power spectrum of the tensor modes ${\mathcal P}(k)$ at the pivot scale $k_*$ can be 
parametrised in terms of the amplitude of the tensor modes $A_T$ and the tensor spectral index $n_T$ as:
\begin{eqnarray}
\label{power}
{\mathcal P}(k) =A_T(k_*) 
\left( \frac{k} {k_*} \right)^{n_T} \,.
\end{eqnarray}
The amplitude $A_T(k_*)$ is related to the amplitude of the scalar modes $A_s (k_*)$ through:
\begin{equation}
A_T(k_*)=A_s(k_*)  r \,,
\end{equation}
where $r$ is the tensor-to-scalar ratio. In the standard slow-roll inflation $n_T$ satisfies the consistency relation 
$n_T \simeq -r/8$, leading to a red-tilted spectrum $n_T<0$.  \\
Alternative inflationary models and particle production models that
allow departures from the consistency relation, leading to a blue-tilted spectrum 
($n_T > 0$), are also discussed in literature \cite{Muko2014,Kuro2021,Kuro2015}.\\
In this analysis we adopt $r = 0.036$, as indicated by the combined {\sc Planck}/BICEP2 observations \cite{Bicep2-Planck}, 
set $A_s(k_*) = 2.0989 \times 10^{-9}$ at $k_* = 0.05,\text{Mpc}^{-1}$ \cite{Planck_par} and take  $n_T$  as  free parameter.

The transfer function $T^2(k)$ is given by:
\begin{equation}
\label{transfer}
T^2(k)=\Omega^2_m \left( \frac{g_{*}(T_{hc})}{g^0_{*}} \right)
\left( \frac{g^0_{*s}}     {g_{*s} (T_{in})} \right)^{4/3}
\left( \frac{3 j_1(z_k)}{z_k}\right)^2 F(k) \,,
\end{equation}
where $g^0_{*}$=3.36 and $g^0_{*s}$=3.91 are the present time effective degrees of freedom for energy density and  for entropy density respectively,
$\Omega_m$=0.31 is the total matter energy density parameter \cite{Planck_par}, $F(k)$ is a fitting function, $j_1(z_k)$ 
is the spherical Bessel function and 
$z_k = k \tau_0$. For $z_k \gg1$, $j_1(z_k)$ can be approximated by 
$j_1(z_k) \simeq 1/ ( \sqrt{2} z_k)$.\\
The horizon crossing temperature $T_{hc}$ associated with the scale $k$, can be expressed as \cite{Nakayama2008}:
\begin{eqnarray}
\label{Thc}
T_{hc}=\frac{k \,M_{pl}} {1.66 \,a_0 \,T_0 \,g^{1/2}_* (T_{hc})} 
 \left( \frac{g_{*s}(T_{hc})} {g^0_{*s}} \right )^{1/3}\,.
\end{eqnarray}
Using $k=a_{hc}H_{hc}$ and applying the entropy conservation
$g_{*s}a^3=$const., the Hubble expansion rate at the horizon crossing $H_{hc}$ is obtained as:
\begin{equation}
\label{Hhc}
H_{hc}=\frac{k^2}{a^2_0 T^2_0 }
\frac{M_{pl}}{1.66} 
\frac{1} {g^{1/2}_* (T_{hc})}
\left(\frac{g_{*s}(T_{hc})}{g^0_{*s}}\right)^{2/3} \,.
\end{equation}
In the above equations "0" and "hc" denote quantities at the present time and at the horizon crossing respectively,
and $T_0\simeq2.725$ K is the present cosmic microwave temperature.\\
The frequency $f$ corresponding to the mode $k$  crossing the horizon at $T_{hc}$ is given by \cite{Saikawa2018}:
\begin{eqnarray}
f=\frac{k} {2 \pi a_0} = 
\frac{H_{hc}} {2 \pi}
\frac{a_{hc}}{a_0} \simeq
2.65 \times 10^{-8}\,
\left(  \frac   {g_{*}    (T_{hc} ) }   {106.75}  \right)^{1/2}
\left(  \frac   {g_{*s}  (T_{hc}) }   {106.75}  \right)^{-1/3}
\left ( \frac{T_{hc}  }  {1 {\rm GeV } }\right){\rm Hz} \,.
\end{eqnarray}

In the standard cosmological model, the fitting function $F(k)$ 
from  Eq. (\ref{transfer}) 
is given by \cite{Nakayama2008,Kuro2015,Turner1993}:
\begin{eqnarray}
F_{st}(k)=T^2_1 \left( \frac{k}{k_{eq}}\right) T^2_2 \left( \frac{k}{k_{RH} }\right) \,
\end{eqnarray}
where the wave numbers corresponding to the matter-radiation equality and the completion of
reheating are respectively:
\begin{eqnarray}
k_{eq} &=& 7.1 \times 10^{-2} {\rm Mpc}^{-1} \Omega_m h^2\,, \\
\label{krh}
k_{RH} & = & 1.7 \times 10^{14} {\rm Mpc}^{-1} \left( \frac{   g_{*s} (T_{RH}) }   {g^0_{*s}} \right)^{1/6} \left(\frac{T_{RH}} {10^7 {\rm GeV} }\right)\,.
\end{eqnarray}
For $T^2_1(x)$, $T^2_2(x)$  we use following fitting functions \cite{Nakayama2008,Kuro2015}:
\begin{eqnarray}
T^2_1(x) &=&1+1.57x+3.42x^2\,,\hspace{0.5cm}
T^2_2(x)=(1-0.22x^{3/2}+0.65x^2)^{-1}\,, \nonumber \\
\end{eqnarray}
In the case with an early matter domination (EMD) phase, $F(k)$ takes the form:
\begin{eqnarray}
F_{EMD}(k)=T^2_1 \left( \frac{k}{k_{eq}}\right)
T^2_2 \left( \frac{k}{k_{dec}}\right)
T^2_3 \left( \frac{k}{k^{S}_{dec} }\right)
T^2_2 \left( \frac{k}{k^{S}_{RH} }\right) \,,
\end{eqnarray}
where:
\begin{eqnarray}
\label{kdec}
k_{dec} & = & 1.7 \times 10^{14}  {\rm Mpc}^{-1} \left( \frac{g_{*s}(T_{dec}) }{ g^0_{*s}}\right)^{1/6}  \left( \frac{T_{dec}}{10^7 {\rm GeV}}\right)\, \\
\label{kdec}
k^S_{dec}&=&k_{dec}\Delta^{2/3}_s\,,\hspace{0.5cm} k^S_{RH}=k_{RH}\Delta^{-1/3}_s\,,
\end{eqnarray}
where $T_{dec}$ denotes the $N_2$ sterile neutrino decay temperature given  in  Eq.  (\ref{Tdec}), $T_{RH}$ represents 
the reheating temperature after inflation and 
$\Delta_s$ is the entropy dilution factor defined in Eq. (\ref{dilut_N2}). \\
For the fitting function $T^2_3(x)$ we use:
\begin{eqnarray}
T^2_3(x) =1+0.59x+0.65x^2\,.
\end{eqnarray}
The characteristic scales $k^S_{dec}$ and $k^S_{RH}$ encode key information about the EMD phase, specifically  its onset,
its end the associated  entropy dilution.

\section{Results}

\subsection{DM signatures in the gravitational waves background}
\begin{figure}
\begin{center}
\includegraphics[width=12cm]{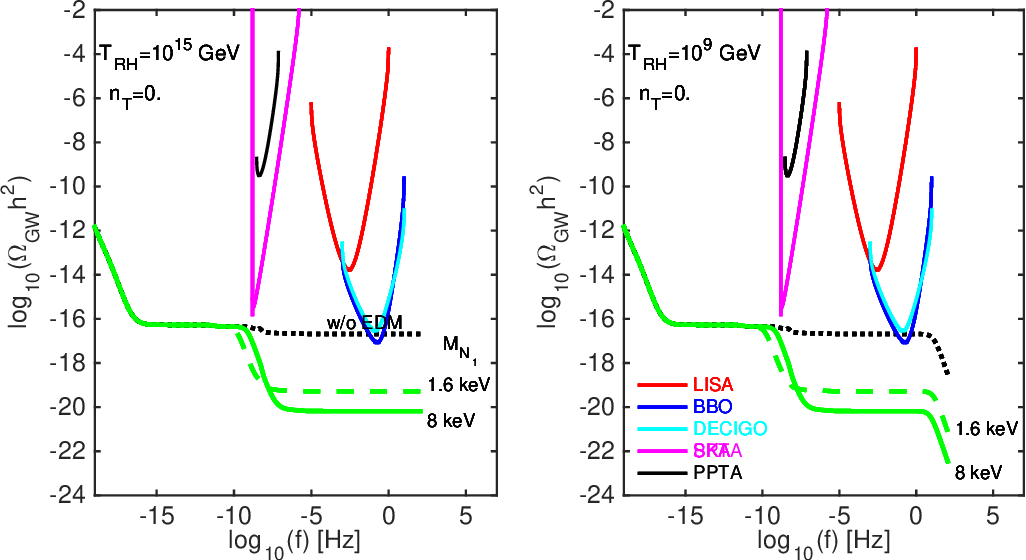}
\caption{ Gravitational wave (GW) energy density spectrum for $n_T=0$. \\
The {\bf left } plot  corresponds to  $T_{RH}=10^{15}$ GeV while the {\bf right} plot presents the results for  $T_{RH}~=~10^9$~GeV. 
In both plots, the GW spectra obtained within the standard model are shown as black dotted lines. The spectra corresponding to scenarios with an early matter domination (EMD) phase are also presented, for the parameter sets $\{M_{N1}(keV)\,,M_{N_2}(GeV)\,,T_{dec}(GeV)\,,\tau_{N_2}(sec.)\}$: \\
 $\{1.6,2.7\,,5.72\times10^{-3}\,,0.224\}$ as green dashed lines, and $\{8\,,80\,,3.1610^{-2}\,, 7.07\times 10^{-3}\}$  as green continuous lines. 
 For comparison, 
the power-law-integrated sensitivity curves (PLISCs) for the future GW experiments \cite{Zendo,Fresh1}  such as SKA, PPTA,
LISA, DECIGO, and BBO are also shown.  \label{fig3}} 
\end{center}
\end{figure}

\begin{figure}
\begin{center}
\includegraphics[width=12cm]{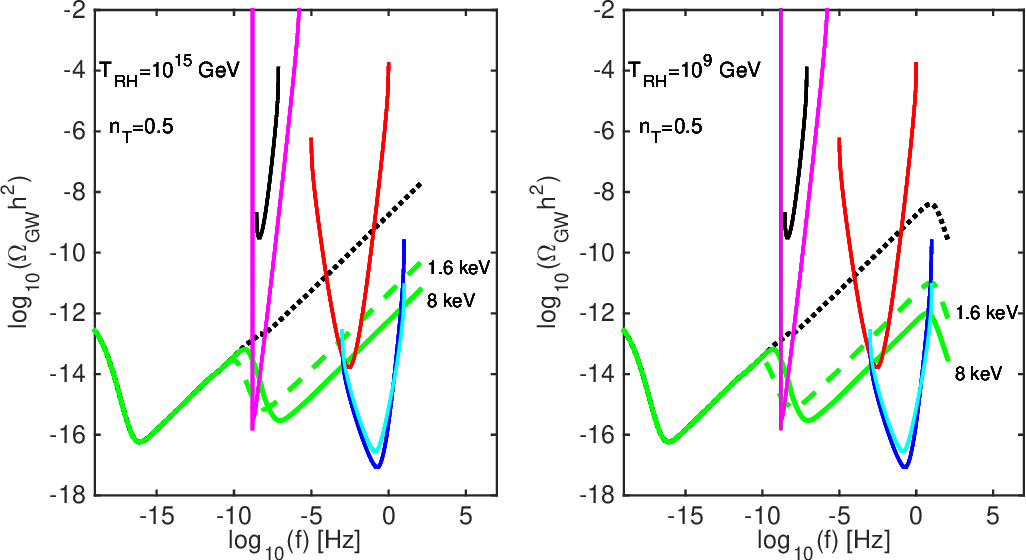}
\caption{The same as in the Figure~\ref{fig3}  for $n_T=0.5$ \label{fig4}} 
\end{center}
\end{figure}

We compute the energy density spectrum of the gravitational waves
  (GWs)  
for the standard model and for models with early matter domination phase (EDM).\\
Figure \ref{fig3} and Figure \ref{fig4} present the GWs energy density spectra obtained 
for tensor spectral index $n_T=0$ and $n_T=0.5$ respectively. \\
Both figures present spectra for two benchmark  reheating temperatures $T_{RH}=10^{15}$ GeV, 
and  $T_{RH}=10^9$GeV.
Each figure shows a characteristic suppression in the GWs spectrum that arises when entropy dilution from $N_2$
decay dominates, characterised by  $k^{S}_{dec}$ given in  Eq. (\ref{kdec}). 
The suppression ends once the $N_2$ decay is completed at $T_{dec}$ and occurs at $k_{dec}$ as defined in Eq. (\ref{kdec}).
This suppression is a direct consequence of the duration of the EMD phase, which is determined by  the $N_2$ neutrino mass $M_{N_2}$ and is lifetime $\tau_{N}$. \\
A larger $\tau_N$  delays the $N_2$ decay, lowering $T_{dec}$  and extending the EMD duration.
These dependencies are clearly illustrated in Figs.~\ref{fig3} and \ref{fig4}, where they are shown by the green dashed lines. \\
In a similar way, increasing $M_{N_2}$ rises $T_{dec}$,  enhancing the entropy dilution and increasing the suppression effect.
These dependencies are shown in Figs.~\ref{fig3} and \ref{fig4} by green continuous lines. \\
The  high frequency suppression for $T_{reh}=10^9$ GeV arises from the inflationary reheating at the characteristic scale 
$k_{RH}$  given in Eq. (\ref{krh}).
Figure \ref{fig3} clearly shows that  for $n_T  = 0$,  
 the prospect of detecting the GW spectrum by the current and upcoming missions
is low. However, a blue-tilted spectrum can enhance the GWs spectrum as illustrated in Figure \ref{fig4} for $n_T = 0.5$.

In the standard inflationary models, the tensor spectral index is typically red, corresponding to a negative tensor tilt ($n_T < 0$). However, a number of non-standard or modified early-universe scenarios allow for a blue tilt tensor spectral index ($n_T > 0$). Examples include super-inflation \cite{Basak2010,Nojiri2022}, phantom inflation \cite{Liu2011,Richarte2017,Iqbal2018}, Galileon inflation \cite{Kobayashi2010,Choudhury2024_NG,Choudhury2024_SIGW}, gas string inflation \cite{Brandenberger2008,Brandenberger2023_review}, as well as multi-field and higher-curvature correction models of inflation \cite{Smith2024,Holland2024}. \\

A blue tilt spectral index $n_T>0$  can enhance the GW spectrum by increasing the effective number of relativistic degrees of freedom
$N_{eff}$ before recombination with its deviation  $\Delta N_{eff}$ from the standatd value $N_{eff}=3.046$ \cite{PDG}.
Thus, the GW spectrum is subject to an upper bound arising from $\Delta N_{eff}$ \cite{Magiore2000}: 
\begin{equation}
\label{intOmegaGW}
\int^{f_{max}}_{f_{min}}  \frac{ {\rm d} f}{f} \Omega_{GW}(f) h^2  \le 5.6 \times 10^{-6} \Delta N_{eff} \,.
\end{equation}
Here $f_{min}$ is the ultraviolet cutoff that is typically $10^{-18}$ Hz for CMB and $10^{-10}$Hz for BBN. 
The ultraviolet cutoff, $f_{max}$, represents the size of the horizon at the end of inflation and depends on the reheating temperature $T_{RH}$ 
\cite{Meerburg2015}. Assuming  instantaneous reheating with  $T_{RH} \sim 10^{15}$ GeV 
one obtains $k_{end} \sim 10^{23}$Mpc$^{-1} $, leading to $f_{max} \simeq 10^8$ Hz \cite{Cabass2016}. \\
Current cosmological observations place upper limits 
on $\Delta N_{eff}$ which in turn constrain the tensor spectral index via Eqs. 
(\ref{OmegaGW} ) and (\ref{intOmegaGW}). 
Specifically, the bounds are $\Delta N_{eff} \le 0.28$ from CMB measurements \cite{Akrami2018} and $\Delta N_{eff} \le 0.4$ from BBN data \cite{Cybrut2016}. \\
Combining the CMB and BBN constraints yields an upper limit on tensor spectral index
$n_T \le 0.4$ (95\% CL) within the standard cosmological model assuming the current 
 tensor-to-scalar ratio $r \le 0.035$ and $T_{RH}\simeq 10^{15}$ GeV \cite{Giare2023}. \\
Direct measurements from ground-based interferometers, such as LIGO and VIRGO, constrain the GWs energy density to 
$\Omega_{GW} \le 10^{-7}$ 
in the frequency range $20 -85.8$ Hz \cite{Abbott2017} which corresponds to an upper limit on the tensor spectral index of 
$n_T \le 0.52$  \cite{Akrami2018}. 

The upper bounds on the tensor spectral index are derived under the assumption 
that the primordial tensor power spectrum follows a power-law behaviour, 
 given in Eq.~(~\ref{power}~), an approach widely adopted in the literature. 
However, the frequencies probed by the GW experiments correspond to modes 
that exited the horizon near the end of inflation. 
Consequently, assuming a pure power-law form across all frequencies can render the estimation of 
$n_T$ unreliable \cite{Giare2021a,Kinney2021}. \\
Ref. \cite{Mel2021} demonstrates that at high frequencies, the primordial tensor power spectrum can exhibit a 
strong dependence on the running of the tensor spectral index
$\alpha_T={\rm d}n_T/{\rm d} ln k$, that parameterise 
the scale-dependence of the tensor tilt. 

The sensitivity of upcoming CMB experiments, such as CMB-S4 \cite{CMB-S4}, CMB-Bharat \cite{CMB-BH}, 
and CMB-HD \cite{CMB-HD}, are expected to significantly improve constraints on the tensor-to-scalar ratio  and $\Delta N_{eff}$.
In particular, the CMB-S4  experiment aims to reach a sensitivity of $r \sim 10^{-3}$ and $\Delta N_{eff} \le 0.05$ that in turn 
will improve the constraints on the tensor tilt.
\begin{figure}
\begin{center}
\includegraphics[width=15cm]{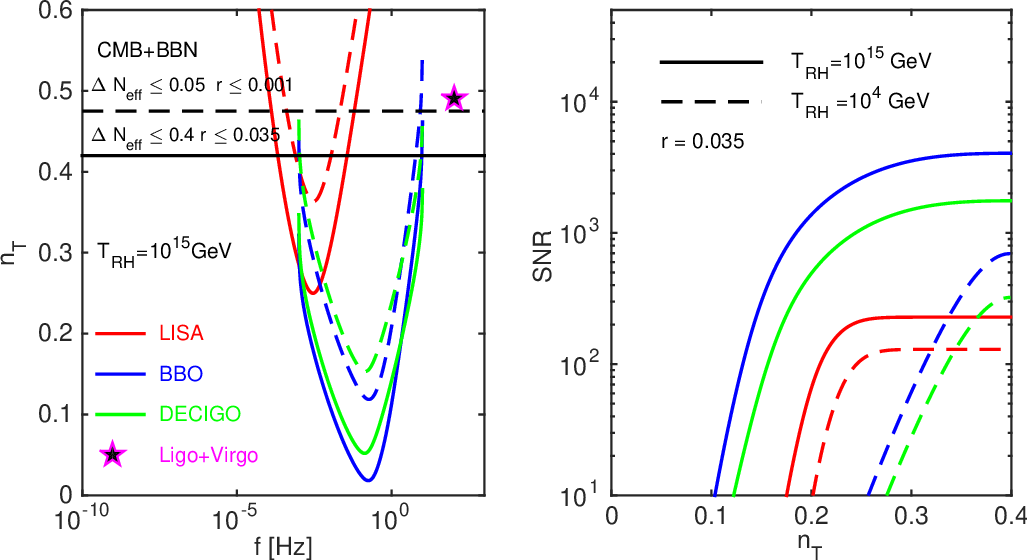} 
\caption{  {\it Left}: Tensor tilt ($n_T$) sensitivity curves for the future space-based interferometer experiments LISA, BBO, and DECIGO, derived from the corresponding power-law-integrated sensitivity curves (PLISCs), are shown for $r \le 0.035$ (solid lines) and $r \le 0.001$ (dashed lines). Also shown are the current and projected upper bounds on $n_T$ from CMB and BBN observations, together with the upper limit on $n_T$ derived from the VIRGO and LIGO experiments. 
{\it Right}: The allowed region in  $(n_T, T_{RH})$ parameter space that yields the signal-to-noise ratio SNR $\ge 10$, obtained for the LISA, BBO, and DECIGO experiments (see also the text). \label{fig5}} 
\end{center}
\end{figure} 
Left panel of Figure~\ref{fig5} shows the tensor spectral index sensitivity curves for the future space-based interferometer experiments LISA \cite{Lisa1,Lisa2}, BBO \cite{BBO1,BBO2}, and DECIGO \cite{DECIGO}, derived from the corresponding 
power-law-integrated sensitivity curves (PLISCs) available online in Zenodo repository \cite{Zendo,Fresh1}.
These results are obtained within the framework 
of standard cosmology for a reheating temperature of $T_{RH} = 10^{15}$ GeV 
and two values of the tensor-to-scalar ratio: the current upper limit $r \le 0.035$ 
and the forecasted value $r \le 0.001$. We also include the current and projected upper 
bounds on $n_T$ from CMB and BBN observations, as well as the tensor tilt constraint from VIRGO and LIGO.

\subsection{Signal-to-noise ratio for GWs experiments} 

To evaluate the capability of various GWs experiments in
searching for the specific early universe evolution imprints in the GWs spectra described above, 
we compute  the signal-to-noise ratio (SNR) 
using the  PLISCs files for the future space interferometers LISA, BBO and DECIGO.
The SNR is defined as:
\begin{eqnarray}
\label{SNR}
SNR \equiv \sqrt{   \tau_{obs} \int^{f_{max}}_{f_{min}}  {\rm d} f \, \left(\frac{\Omega_{GW}^{EDM}(f)h^2}{\Omega^{exp}_{GW}(f)h^2} \right)^2    }\,,
\end{eqnarray}
where $\Omega^{EDM}_{GW}(f)$ are the GWs energy density spectra discussed in the previous section,
 $\Omega^{exp}_{GW}(f)$ denotes the energy density spectra derived from the PLICs signals,
 $f_{min}$ and $f_{max}$ define the operational frequency range corresponding to different experiments and 
 $\tau_{obs}$ is the total observation time.
 For a consistent sensitivity comparison, we adopt a common observation time of
 $\tau_{obs}~=~5$~years for all experiments and impose the detection threshold at $SNR \ge 10$ \cite{Caprini2018}. 
 
  In this analysis, we treat $n_T$ and $T_{RH}$ as free parameters within the region of parameter space allowed 
  by CMB and BBN constraints for $r \le 0.035$, starting from  $T_{RH}=10^{15}$ GeV.
  The lower limit of $T_{RH} = 10^{4}$ GeV is imposed by the 
  requirement that the reheating temperature must exceed the mass of the heaviest right-handed (RH) neutrino, $M_{N_3}$.
The right panel of Figure~\ref{fig5} shows the allowed region   $(n_T, T_{RH})$  that yields the signal-to-noise ratio SNR $\ge 10$, 
obtained for the LISA, BBO, and DECIGO experiments.
The high-frequency suppression of $\Omega_{GW}^{EDM}(f)$ resulting from inflationary reheating at 
the characteristic scale $k_{RH}$  is manifested as a decrease in the signal-to-noise ratio with decreasing $T_{RH}$.
Since the SNR scales with $T_{RH}$, for clarity, the figure presents the results obtained for 
for  $T_{RH} = 10^{15}$ GeV and $T_{RH} = 10^{4}$ GeV.

\section{Conclusions}

In this work, we consider the left-right symmetric extension of the Standard Model (LRSM), which introduces a right-handed charged gauge boson, 
$W_R$, with a mass at the TeV scale. This framework naturally accommodates the type-II seesaw mechanism for generating active neutrino masses. \\
The right-handed $W_R$ boson plays a central role in determining the dark matter relic abundance. 
For RH neutrinos that decouple while still relativistic, it ensures, to very good approximation, that their freeze-out temperatures, and thus their yields, are identical. \\
At later times, once the heavier RH neutrinos become non-relativistic, they behave as matter and come to dominate the energy density of the universe. 
This leads to a period of early matter domination (EMD), which ends when the RH neutrinos decay, releasing a significant amount of entropy.

We analyse the conditions required to achieve the appropriate entropy dilution that 
both aligns the sterile neutrino dark matter abundance with observational constraints 
and induces an early matter domination phase. 
The latter leaves a characteristic, frequency-dependent suppression in 
the spectral shape of the stochastic gravitational wave background. 
Furthermore, we find that these conditions impose bounds on  the lightest 
active neutrino  mass: $8.59 \times 10^{-10} {\rm eV} < m_{\nu_1} < 5.06 \times 10^{-9} {\rm eV}$.

We demonstrate that the frequency-dependent suppression of the gravitational wave background arises 
directly from the duration of the early matter domination  phase, which is governed by the mass and lifetime of the heavier RH neutrino. 
To assess the detectability of this suppression in the GW energy spectrum by the 
upcoming experiments such as SKA, LISA, BBO, and DECIGO, we evaluate the corresponding signal-to-noise ratio (SNR). \\
We find that a blue-tilted primordial tensor power spectrum can significantly enhance the GW energy density, enabling the detection 
with $SNR > 10$ in experiments such as LISA, BBO, and DECIGO. 

The main challenge faced in this model is the mass of right-handed gauge boson  
 $M_{WR} \sim 10$ TeV. Theoretical considerations  set  a lower bound  $M_{WR} >2.5-4$ TeV \cite{MW1,Neme2012}, 
 while direct searches as ATLAS and CMS continue to raise the experimental limits on $M_{W_R}$ \cite{PDG}. 
 Alternative approaches to avoid the overproduction of sterile neutrino dark matter within the LRSM
  have been proposed. For instance, Ref. \cite{Neme2012} discusses a scenario in which all new gauge interactions are realised at the QCD scale. 

\section{Acknowledgment}
I would like to thank  to Alexandru Dobrin for useful comments on the manuscript.
The author acknowledge the use of the computing facilities at the Institute of Space Science.

\end{document}